\documentclass[11pt,a4paper]{article}

\usepackage{listings}
\usepackage{array}
\usepackage{tabularx}
\usepackage[inline]{enumitem}
\usepackage{xltabular}
\usepackage{booktabs} 
\usepackage{graphicx} 
\usepackage{ragged2e} 
\usepackage{changepage} 
\usepackage[top=1in, bottom=1in, left=1in, right=1in]{geometry}
\usepackage{caption}

\usepackage{authblk}

\title{\textbf{Agentic AI and Pedagogical Best Practice: The Tension Between Automation and Learning}}

\author[1]{Steve Woollaston\thanks{Corresponding author: s.m.woollaston@gmail.com}}
\author[2]{Brendan Flanagan}
\author[3]{Isanka Wijerathne}
\author[3]{Hiroaki Ogata}

\affil[1]{Graduate School of Informatics, Kyoto University, Japan}
\affil[2]{College of Information Science and Engineering, Ritsumeikan University, Japan}
\affil[3]{Academic Center for Computing and Media Studies, Kyoto University, Japan}

\date{} 

\usepackage{fancyhdr}
\begin{document}

\maketitle
\thispagestyle{fancy} 

\begin{abstract}
Artificial intelligence in education is evolving from passive chatbots to proactive AI agents capable of initiation and goal-directed interactions. While offering opportunities for personalised learning, this shift risks undermining learner agency and cognitive effort. This paper reviews six pedagogical principles—prior knowledge activation, collaborative learning, problem-based learning, formative assessment, scaffolding, and metacognition—through the lens of agentic AI. We discuss the tension between automation and learning, proposing design recommendations that prioritise intentional friction, dynamic scaffolding, human-in-the-loop oversight, and considered AI utilisation to ensure AI supports rather than supplants human learning.
\end{abstract}


\clearpage

\section{Introduction}

The landscape of artificial intelligence in education is experiencing another paradigm shift, evolving from passive, reactive chatbots into goal-directed, proactive AI agents \cite{Deng2025-yq}. Fundamentally, agentic systems function as digital actors that can independently monitor their surroundings, process and filter information to align with specific objectives, and carry out purposeful tasks to reach those targets \cite{gulli2025agentic}. Unlike conventional dialogue systems that merely follow user instructions, proactive AI agents possess the ability to anticipate outcomes, take initiative, and strategically plan to achieve specific goals \cite{Acharya2025-pv}. While this transition offers numerous opportunities for personalised and realtime adaptive learning for each student \cite{Martinez2025-ly}, it also introduces a challenge: designing autonomous systems that support learning without undermining learner agency \cite{Mouta2025-wg} or the professional judgment of teachers \cite{Petersen2019-jp}.

Agentic systems, operating within the broader digital ecosystem of an educational application, consist of three core behavioural traits: \textbf{autonomy} to function without continuous human intervention, \textbf{proactiveness} to initiate goal-directed actions, and \textbf{reactiveness} to adapt to changing contexts (e.g., input from users, other agents, or the digital environment itself) \cite{gulli2025agentic}. A key functional advantage of this setup is the system's ability to utilise external tools, such as databases, APIs, and microservices, while leveraging persistent memory to maintain continuity across sessions and communicate across multi-agent networks \cite{gulli2025agentic}.

As educational technology becomes more sophisticated, there is a risk that AI automation bypasses the cognitive effort necessary for deep learning. While productive cognitive offloading (delegation of routine mental tasks to external tools) can potentially free up mental bandwidth for higher-order reasoning, it frequently shifts into cognitive surrender. In this state, the learner effectively abdicates intellectual agency, allowing the AI to perform the critical synthesis and analysis that the student should be doing themselves for learning to occur \cite{Chirayath2025-bd, Shaw2026-ne}. Research cautions that an over-reliance on technology can undermine essential human elements of teaching and learning, such as deep processing, focus, critical thinking, empathy, resilience, and creativity \cite{Martinez2025-ly}. To avoid this, agentic AI systems must be intentionally designed to promote core pedagogical principles rather than replace learner cognition. Successful learning with agentic AI requires deliberate instructional \textit{friction} and the strategic \textit{fading} of AI support as students build competence. By balancing proactive scaffolding with necessary and beneficial cognitive challenges, AI can preserve the productive struggle essential to effective learning \cite{kapur2015learning}.

To anchor the design of these agentic AI systems in educational best practice, this paper presents and discusses established findings from educational literature on learning, effective instruction, and pedagogical theory \cite{Husbands2012-qd, Rosenshine2012-zt} in terms of agentic AI systems. Six best practice principles were chosen to examine more closely, and then discussed in terms of the opportunities and challenges agentic AI systems provide. In no particular order, these are presented below.

\section{Pedagogical Principles and Agentic AI}

\subsection{Prior Knowledge \& Experience}

Activating prior knowledge is foundational to learning; instruction should begin with a short review of previous material to strengthen fluent recall and help learners integrate new information into existing cognitive schemas \cite{James2014-so}. Effective pedagogy should also recognise and build upon what learners already know by taking account of the personal and cultural experiences of different groups \cite{Chan2019-xl}.

\textbf{Opportunity - Context-Aware Retrieval \& Personalisation:} Agentic AI offers opportunities to facilitate this process through context-aware retrieval. By leveraging persistent learner profiles, an AI tutor can proactively bridge new concepts to a student's specific past lessons or personal interests. For example, framing a new mathematical concept through the lens of a student's known hobby reduces cognitive load and frees up working memory for deeper processing.

\textbf{Challenge - Algorithmic Assumptions \& Cultural Misalignment:} Use of agentic AI introduces the risk of algorithmic bias and cultural misalignment. AI systems may make flawed inferences about a student's prior knowledge based on algorithmic generalisations, incomplete and / or biased training data. Furthermore, AI agents often lack the nuanced local cultural competence of a human teacher; attempting to relate a concept to a student's informal background could result in stereotypical, irrelevant, or tone-deaf connections that alienate the learner rather than engage them.

\subsection{Collaborative \& Team-Based Learning}

Collaborative learning encompasses instructional methods where students work together in small groups toward a collective goal, relying heavily on positive interdependence, individual accountability, and social skills \cite{Dzaiy2024-hv}. This approach fosters vital interpersonal skills, teaching students how to articulate reasoning, listen to differing ideas without defensiveness, and navigate compromise \cite{Fink2016-xn}.

\textbf{Opportunity - Proactive Teammate:} Large language models (LLM) thrive in roleplays and can act as a proactive teammate. A conversational AI agent could simulate a team member to prompt counter-arguments as a ``devil's advocate'' or monitor group dynamics to encourage equitable participation and co-regulation.

\textbf{Challenge - Loss of Authentic Connection:} Integrating AI into peer learning poses a critical challenge regarding the loss of authentic connection, which is built on trust, respect, and empathy. There is a profound risk that an overly assertive AI agent might dominate group work or resolve disputes too efficiently. If AI preempts the messy but necessary process of peer negotiation, it threatens to stunt the development of genuine human empathy, conflict resolution, and the deep interpersonal skills that make collaborative learning so valuable.

\subsection{Problem-Based Learning}

Problem-Based Learning (PBL) centres on engaging students with authentic, ill-structured, real-world problems that require self-directed research and collective brainstorming \cite{Kim2011-lw}. By grappling with complex scenarios, students develop research skills, critical thinking, grit, self-regulated learning, and interdisciplinary problem-solving skills.

\textbf{Opportunity - Dynamic Simulation:} Agentic AI can generate infinitely customisable, context-rich scenarios. An AI system can act as a realistic stakeholder, client, or interactive environment that students must interview and satisfy, bringing realism and adaptability to the learning space.

\textbf{Challenge - Undermining Productive Struggle:} This capability introduces the challenge of undermining productive struggle. The pedagogical value of PBL relies heavily on students grappling with ambiguity and independently navigating the inquiry process \cite{Khan2020-gq}. An overly ``helpful'' AI might preemptively solve the complex problem or prematurely reduce the cognitive load, stripping away the necessary instructional friction. If the AI resolves the ambiguity too quickly, the fundamental pedagogical value of the struggle is lost, transforming a robust problem-solving exercise into the passive receipt of information.

\subsection{Formative Assessment and Realtime Feedback}

Effective pedagogy embeds continuous, formative assessment during the learning process to provide immediate corrective feedback and identify comprehension gaps \cite{Hernandez-de-Menendez2019-kv}. By checking for understanding frequently, educators can determine how well material and skills have been learned, adapt teaching strategies in real-time, and ensure content and learning activities are within the student's zone of proximal development (ZPD) \cite{Rosenshine2012-zt}. Research consistently highlights that high-quality feedback is one of the most powerful influences on student achievement \cite{Husbands2012-qd}.

\textbf{Opportunity - Process vs.\ Product:} Agentic AI presents an opportunity to move further beyond outcome evaluation toward continuous, unobtrusive process-based assessment. By tracking how a student solves a problem step-by-step, AI can highlight specific cognitive misconceptions and deliver personalised, just-in-time feedback without waiting for summative assessments.

\textbf{Challenge - Assessment Integrity \& Privacy:} However, continuous monitoring introduces significant ethical implications regarding privacy and surveillance. Furthermore, as AI becomes deeply integrated into the workflow, it becomes increasingly difficult to disentangle authentic student capability from AI assistance, challenging the validity of the assessment itself.

\subsection{Scaffolding}

Scaffolding involves providing temporary cognitive, social, and emotional support, such as step-by-step demonstrations, expert models, or prompts, to help learners navigate difficult tasks \cite{Puntambekar2003-jp}. These supports are designed to be transitional; as the learner's competence and understanding increase, the scaffolds are gradually withdrawn or ``faded'' to foster independent problem-solving \cite{Rosenshine2012-zt}.

\textbf{Opportunity - Personalised \& Teachable AI:} Agentic AI can offer 24/7 personalised tutoring that patiently breaks down complex tasks and explicitly models expert thinking for individual students. Alternatively, it can act as ``teachable AI,'' where the system simulates a novice that the student instructs, leveraging Māori \textit{tuakana-teina} pedagogy (teacher and student reciprocally learning from each other) to deepen the student's own mastery of the subject \cite{Woods2011-tt}.

\textbf{Challenge - The ``Fading'' Problem:} The core challenge lies in the ``fading'' problem, a difficulty shared by both human educators and AI. Generative AI inherently provides implicit scaffolding through its adaptive, iterative nature, making it hard to gauge when and how to withdraw support without hindering the learner. Over-reliance on this automated assistance risks inducing ``learned helplessness'' which undermines the goal of student autonomy \cite{maier1976learned}. Research suggests that a primary focus should therefore shift from simply managing the removal of support to fostering the student's capacity for reflective, self-regulated interaction with the AI \cite{chrysanthi2026scaffolding}.

\subsection{Promoting Metacognition \& Reflection}

Promoting metacognition requires teaching self-regulatory skills and prompting students to consciously analyse their own learning processes, goals, and emotions \cite{Lemberger2012-yk}. Deliberately focussing attention on these allows students to ``think about their thinking,'' evaluate their strategies, and adaptively integrate new knowledge. As Dewey \cite{Dewey1933-qd} asserts, learning does not happen merely from experience, but from actively reflecting on that experience.

\textbf{Opportunity - Reflective Prompts:} Agentic AI can facilitate this self-regulation by proactively interrupting learning sequences at optimal moments to ask metacognitive questions, such as \textit{``Why did you choose this approach?''} or \textit{``What is your goal here?''} This encourages continuous meta-learning and reflection without requiring a teacher to manually prompt each student.

\textbf{Challenge - Superficial Compliance vs.\ Internal Regulation:} The critical challenge lies in designing AI interventions that provoke genuine contemplation rather than shallow compliance. Students often learn to ``game'' automated systems, providing superficial answers merely to bypass the AI's reflective checkpoints. Furthermore, if an AI over-manages the reflection process, it risks permanently externalising a skill that must ultimately become an internal, self-directed habit.

\section{Agentic AI Implementation Matrix}

To bridge theory, system architecture, and practice, Table~\ref{tab:implementation_matrix} operationalises these principles into an actionable design framework. This matrix maps each pedagogical concept to specific system actions, demonstrating how the strategies of intentional instructional friction and dynamic scaffolding fading can be embedded across diverse learning contexts.
\renewcommand{\arraystretch}{1.5}
\small

\begin{adjustwidth}{-1.25cm}{-1.25cm}

\begin{xltabular}{\textwidth}{%
  >{\RaggedRight\arraybackslash}p{0.18\textwidth} 
  >{\RaggedRight\arraybackslash}X 
  >{\RaggedRight\arraybackslash}X 
  >{\RaggedRight\arraybackslash}X}
  
\caption{Pedagogically Informed Agentic AI Implementation Matrix}\label{tab:implementation_matrix} \\

\toprule
\multicolumn{1}{c}{\textbf{Principle}} & 
\multicolumn{1}{c}{\textbf{Agent Action}} & 
\multicolumn{1}{c}{\textbf{Friction / Fading}} & 
\multicolumn{1}{c}{\textbf{Learning Example}} \\ 
\midrule
\endfirsthead

\toprule
\multicolumn{1}{c}{\textbf{Principle}} & 
\multicolumn{1}{c}{\textbf{Agent Action}} & 
\multicolumn{1}{c}{\textbf{Friction / Fading}} & 
\multicolumn{1}{c}{\textbf{Learning Example}} \\ 
\midrule
\endhead

\textbf{Prior Knowledge \& Experience} & Agent(s) scan student logs and profile to pull past learning contexts or explicit interests (e.g., hobbies, local geography) to introduce a new concept. & \textbf{Friction:} The system asks the student to explain \textit{how} their hobby relates to the new topic, rather than doing the explaining for them. & A student struggling with coordinates is prompted: \textit{``Remember when you mapped out the field for touch rugby last term? How would you describe your position using a grid?''} \\ \addlinespace

\textbf{Collaborative \& Team-Based Learning} & AI acts as a structured ``co-pilot'' or ``devil's advocate'' for a small group, feeding prompts to ensure equitable turn-taking. & \textbf{Fading:} The AI explicitly scales back its prompts as the group establishes a healthy conversational rhythm (\textit{co-regulation}). & In a history group discussion, the AI messages the tablet: \textit{``Group, ask John what he thinks about the second source before moving on.''} \\ \addlinespace

\textbf{Problem-Based Learning (PBL)} & System generates an interactive simulation of a complex, local, real-world issue, acting as an adaptive stakeholder. & \textbf{Friction:} The AI refuses to provide a direct ``solution pathway'' when stuck, offering clues about \textit{where to look} instead. & Students tackle a local river pollution problem. The AI acts as a local farmer or regional councillor whom students must interview to gather data. \\ \addlinespace

\textbf{Formative Assessment \& Realtime Feedback} & Tracks step-by-step problem-solving (e.g., lines of code, math working steps) and diagnoses misconceptions mid-process. & \textbf{Fading:} Feedback shifts from explicit corrections to subtle hints, and eventually to zero cues as mastery increases. & A math student misapplies a formula. The AI highlights the specific line: \textit{``Look closely at this step. What happened to the negative sign here?''} \\ \addlinespace

\textbf{Scaffolding} & Breaks large tasks into micro-steps or acts as a ``novice'' that the student must teach. & \textbf{Fading:} The AI dynamically removes intermediate step prompts based on the student's speed and accuracy over past tasks. & A student writes an essay. The AI initially provides paragraph structures, but for the final paragraphs, it removes the template entirely. \\ \addlinespace

\textbf{Promoting Metacognition} & Integrates timed, micro-pauses in the workflow, requiring students to evaluate their strategy or emotional state. & \textbf{Friction:} System detects rapid clicking or superficial answers and pauses the interface, prompting a verbal or typed rationale. & After a reading task, the AI prompts: \textit{``You changed your answer twice on question 3. What made you doubt your first instinct?''} \\ 
\bottomrule

\end{xltabular}

\end{adjustwidth}


\section{Discussion: Human Learning and AI Agency}

The integration of proactive AI agents into educational settings presents opportunities and challenges. Across the principles discussed a recurring tension emerges between \textbf{automation} and \textbf{learning}. While AI's ability to reduce cognitive load and provide instant scaffolding is technically impressive, it creates a paradoxical risk: by optimising the efficiency of task completion, we may inadvertently bypass the cognitive effort required for deep, meaningful learning. True agency requires the learner to grapple with complexity, a process that cannot be fully outsourced to an algorithm without stripping away the value of the educational experience.

To ensure these technologies support rather than supplant human development, we propose these design recommendations:

\begin{itemize}
    \item \textbf{Design for Friction:} Rather than aiming for seamless, friction-free interfaces, AI systems should be intentionally designed to withhold answers and support. By engineering strategic ``instructional friction,'' agents can force the critical thinking and active inquiry necessary to cement knowledge, ensuring the learner remains the primary driver of the intellectual journey.
    \item \textbf{Implement Dynamic Fading:} AI scaffolding must be inherently transitional, explicitly designed to retreat as student competence increases. To prevent learned helplessness, agentic systems require robust, real-time analytics capable of distinguishing between temporary task completion and long-term mastery, allowing the AI to dynamically withdraw intermediate supports and systematically yield full cognitive autonomy back to the learner \cite{taber2018scaffolding}.
    \item \textbf{Architectural ``Teacher-in-the-Loop'':} Unlike standard generative AI systems that relegate educators to passive observers of student-chatbot interactions, an agentic educational architecture must include the teacher as an active coordinator of the agent's state machine. This requires three distinct mechanisms: 
    \begin{enumerate}[label=(\arabic*), leftmargin=2em]
        \item \textit{Escalation Protocols}: Where the agent automatically pauses its execution loop and yields control to the teacher when encountering predefined issues or instructional thresholds, such as persistent student misconception loops, excessive frustration or misuse, or content and skills which require relational teaching or special sensitivity;
        \item \textit{Purpose and Guardrail Adjustability}: Allowing teachers to adjust each agent's goals, prompt boundaries, tool access, and autonomy levels in real-time based on live classroom dynamics; and
        \item \textit{State-Interruptibility}: Giving educators the power to manually override and adjust an agent's planned trajectory mid-session.
    \end{enumerate}
    By embedding these interventions directly into the agent's execution cycle, the system ensures that the human educator remains genuinely empowered as a pedagogical director rather than a bystander to automation.

    \item \textbf{AI Usage Restraint:} \textit{Just because we can, doesn't always mean we should.} Educators should critically assess which learning tasks genuinely benefit from AI support and which are better served through traditional methods that prioritise human interaction and internal cognitive processing. Applying the SAMR model \cite{Hamilton2016-gd}, we should avoid the use of AI merely for \textit{Substitution} or \textit{Augmentation}, where the technology adds little pedagogical value. Instead, we should prioritise instances where AI can facilitate true \textit{Modification} or \textit{Redefinition} of the learning experience. Sometimes, the most effective tool for deep learning remains a conversation or a blank page.
\end{itemize}

We must prioritise pedagogical integrity over efficiency. The focus should remain on the promotion of human learning, even when the convenience of automation is available. We urge educators and AI stakeholders to consider Sternberg's \cite{Sternberg2024-uz} warning carefully: \textbf{Ask not what AI can do for you; ask what AI is doing to you.} Our goal should be to create tools and processes that strengthen each student's capacity to think, feel, and create, rather than tools that act as a substitute for the hard, necessary work of thinking.

\section*{Acknowledgments}I am deeply grateful to Lyn MacKenzie for her invaluable mentorship and wisdom during my time as a school teacher, which have been instrumental in growing my practice and refining my pedagogy. This research is supported by Council for Science, 3rd SIP JPJ012347, and JSPS Grant-in-Aid for Scientific Research (B) JP23H01001, (A) JP23H00505, and KAKENHI Grant Number 26KJ1488. 

\section*{Declaration on Generative AI}

 During the preparation of this work, the author(s) used Gemini 3.1 in order to: check grammar and spelling, paraphrase and reword sections for clarity, and assist in drafting the abstract. After using this tool, the author(s) reviewed and edited the content as needed and take full responsibility for the publication’s content.

\bibliographystyle{plain}
\bibliography{hai_aied_references}

@ARTICLE{Acharya2025-pv,
  title        = {Agentic {AI}: Autonomous intelligence for complex goals—A comprehensive survey},
  author       = {Acharya, Deepak Bhaskar and Kuppan, Karthigeyan and Divya, B},
  journal      = {IEEE Access: Practical Innovations, Open Solutions},
  publisher    = {Institute of Electrical and Electronics Engineers (IEEE)},
  volume       = {13},
  pages        = {18912--18936},
  year         = {2025},
  doi          = {10.1109/access.2025.3532853}
}

@ARTICLE{Rosenshine2012-zt,
  title        = {Principles of instruction: Research-based strategies that all teachers should know},
  author       = {Rosenshine, Barak},
  journal      = {American Educator},
  publisher    = {American Federation of Teachers},
  volume       = {36},
  number       = {1},
  pages        = {12},
  year         = {2012},
  language     = {en}
}

@ARTICLE{Kim2011-lw,
  title        = {Scaffolding problem solving in technology-enhanced learning environments ({TELEs}): Bridging research and theory with practice},
  author       = {Kim, Minchi C and Hannafin, Michael J},
  journal      = {Computers \& Education},
  publisher    = {Elsevier BV},
  volume       = {56},
  number       = {2},
  pages        = {403--417},
  year         = {2011},
  doi          = {10.1016/j.compedu.2010.08.024},
  language     = {en}
}

@ARTICLE{Dzaiy2024-hv,
  title        = {The use of active learning strategies to foster effective teaching in higher education institutions},
  author       = {Dzaiy, Aryan Hussein Sulaiman and Abdullah, Saman Ahmed},
  journal      = {Zanco for the Humanities},
  publisher    = {Salahaddin University - Erbil},
  volume       = {28},
  number       = {4},
  pages        = {328--351},
  year         = {2024},
  doi          = {10.21271/zjhs.28.4.18}
}

@phdthesis{Khan2020-gq,
  title     = {{Epistemic Network Analysis} in {Problem-Based Learning} ({PBL})},
  author    = {Khan, Mahmud Hasan},
  school    = {University of Eastern Finland},
  publisher = {University of Eastern Finland},
  year      = {2020},
  doi = {10.13140/RG.2.2.14315.36643}
}

@ARTICLE{Puntambekar2003-jp,
  title        = {Distributed scaffolding: Helping students learn science from design},
  author       = {Puntambekar, S and Kolodner, J L},
  journal      = {Cognition and Instruction},
  volume       = {21},
  number       = {1},
  pages        = {1--47},
  year         = {2003},
  doi          = {10.1207/S1532690XCI2101_01}
}

@ARTICLE{Chan2019-xl,
  title        = {Te whāriki: An early childhood curriculum in a superdiverse {New Zealand}},
  author       = {Chan, Angel},
  journal      = {New Zealand Journal of Educational Studies},
  publisher    = {Springer Science and Business Media LLC},
  volume       = {54},
  number       = {2},
  pages        = {245--259},
  year         = {2019},
  doi          = {10.1007/s40841-019-00138-z},
  language     = {en}
}

@ARTICLE{Sternberg2024-uz,
  title        = {Do not worry that generative {AI} may compromise human creativity or intelligence in the future: It already has},
  author       = {Sternberg, Robert J},
  journal      = {Journal of Intelligence},
  publisher    = {MDPI AG},
  volume       = {12},
  number       = {7},
  pages        = {69},
  year         = {2024},
  doi          = {10.3390/jintelligence12070069},
  language     = {en}
}

@TECHREPORT{Husbands2012-qd,
  author      = {Husbands, Chris and Pearce, Jo},
  title       = {Great pedagogy: Nine claims from research},
  institution = {National College for School Leadership},
  year        = {2012},
  type        = {Research Report},
  language    = {en}
}

@ARTICLE{Hernandez-de-Menendez2019-kv,
  title        = {Active learning in engineering education. A review of fundamentals, best practices and experiences},
  author       = {Hern{\'e}ndez-de-Men{\'e}ndez, Marcela and Vallejo Guevara, Antonio and Tud{\'o}n Mart{\'i}nez, Juan Carlos and Hern{\'e}ndez Alc{\'a}ntara, Diana and Morales-Menendez, Ruben},
  journal      = {International Journal on Interactive Design and Manufacturing (IJIDeM)},
  publisher    = {Springer Science and Business Media LLC},
  volume       = {13},
  number       = {3},
  pages        = {909--922},
  year         = {2019},
  doi          = {10.1007/s12008-019-00557-8},
  language     = {en}
}

@ARTICLE{Lemberger2012-yk,
  title        = {Student Success Skills: An evidence-based cognitive and social change theory for student achievement},
  author       = {Lemberger, Matthew E and Brigman, Greg and Webb, Linda and Moore, Molly M},
  journal      = {Journal of Education},
  publisher    = {SAGE Publications},
  volume       = {192},
  number       = {2-3},
  pages        = {89--99},
  year         = {2012},
  doi          = {10.1177/0022057412192002-311},
  language     = {en}
}

@BOOK{Dewey1933-qd,
  title     = {How we think: A restatement of reflective thinking to the educative process},
  author    = {Dewey, J},
  publisher = {D.C. Heath and Company},
  year      = {1933}
}

@ARTICLE{Woods2011-tt,
  title        = {Reflections on pedagogy: A journey of collaboration},
  author       = {Woods, Christine},
  journal      = {Journal of Management Education},
  publisher    = {SAGE Publications},
  volume       = {35},
  number       = {1},
  pages        = {154--167},
  year         = {2011},
  doi          = {10.1177/1052562910384936},
  language     = {en}
}

@ARTICLE{Hamilton2016-gd,
  title        = {The substitution augmentation modification redefinition ({SAMR}) model: A critical review and suggestions for its use},
  author       = {Hamilton, Erica R and Rosenberg, Joshua M and Akcaoglu, Mete},
  journal      = {TechTrends: For Leaders in Education \& Training},
  publisher    = {Springer Science and Business Media LLC},
  volume       = {60},
  number       = {5},
  pages        = {433--441},
  year         = {2016},
  doi          = {10.1007/s11528-016-0091-y},
  language     = {en}
}

@ARTICLE{Martinez2025-ly,
  title        = {Active learning strategies: A mini review of evidence-based approaches},
  author       = {Martinez, Maria Eugenia and Gomez, Valeria},
  journal      = {Acta Pedagogia Asiana},
  publisher    = {Tecno Scientifica Publishing},
  volume       = {4},
  number       = {1},
  pages        = {43--54},
  year         = {2025},
  doi          = {10.53623/apga.v4i1.555},
  language     = {en}
}

@ARTICLE{Chirayath2025-bd,
  title        = {Cognitive offloading or cognitive overload? How {AI} alters the mental architecture of coping},
  author       = {Chirayath, Ginto and Premamalini, K and Joseph, Jeena},
  journal      = {Frontiers in Psychology},
  publisher    = {Frontiers Media SA},
  volume       = {16},
  pages        = {1699320},
  year         = {2025},
  doi          = {10.3389/fpsyg.2025.1699320},
  language     = {en}
}

@ARTICLE{Deng2025-yq,
  title        = {Proactive conversational {AI}: A comprehensive survey of advancements and opportunities},
  author       = {Deng, Yang and Liao, Lizi and Lei, Wenqiang and Yang, Grace Hui and Lam, Wai and Chua, Tat-Seng},
  journal      = {ACM Transactions on Information Systems},
  publisher    = {Association for Computing Machinery (ACM)},
  volume       = {43},
  number       = {3},
  pages        = {1--45},
  year         = {2025},
  doi          = {10.1145/3715097},
  language     = {en}
}

@ARTICLE{Shaw2026-ne,
  title        = {Thinking-fast, slow, and artificial: How {AI} is reshaping human reasoning and the rise of cognitive surrender},
  author       = {Shaw, Steven D and Nave, Gideon},
  journal      = {Available at SSRN 6097646},
  year         = {2026}
}

@INBOOK{James2014-so,
  title     = {{TLRP’s} ten principles for effective pedagogy: rationale, development, evidence, argument and impact},
  author    = {James, Mary and Pollard, Andrew},
  booktitle = {Principles for Effective Pedagogy},
  publisher = {Routledge},
  pages     = {14--67},
  year      = {2014},
  doi       = {10.4324/9781315872643-6}
}

@ARTICLE{Mouta2025-wg,
  title        = {“where is agency moving to?”: Exploring the interplay between {AI} technologies in education and human agency},
  author       = {Mouta, Ana and Pinto-Llorente, Ana Mar{\'i}a and Torrecilla-S{\'a}nchez, Eva Mar{\'i}a},
  journal      = {Digital Society: Ethics, Socio-Legal and Governance of Digital Technology},
  publisher    = {Springer Science and Business Media LLC},
  volume       = {4},
  number       = {2},
  year         = {2025},
  doi          = {10.1007/s44206-025-00203-9},
  language     = {en}
}

@ARTICLE{Petersen2019-jp,
  title        = {Preservice student views of teacher judgement and practice in the age of artificial intelligence},
  author       = {Petersen, N and Batchelor, J},
  journal      = {Southern African Review of Education},
  volume       = {25},
  number       = {1},
  pages        = {70--88},
  year         = {2019},
  doi          = {10.10520/ejc-1877d530c6}
}

@ARTICLE{Fink2016-xn,
  title        = {Five high-impact teaching practices: A list of possibilities},
  author       = {Fink, L Dee},
  journal      = {Collected Essays on Learning and Teaching},
  publisher    = {University of Windsor Leddy Library},
  volume       = {9},
  pages        = {3},
  year         = {2016},
  doi          = {10.22329/celt.v9i0.4428},
  language     = {en}
}

@BOOK{gulli2025agentic,
  title     = {Agentic design patterns: a hands-on guide to building intelligent systems},
  author    = {Gull{\'\i}, Antonio},
  year      = {2025},
  publisher = {Springer Nature},
  doi       = {10.1007/978-3-032-01402-3}
}

@INCOLLECTION{taber2018scaffolding,
  title     = {Scaffolding learning: Principles for effective teaching and the design of classroom resources},
  author    = {Taber, Keith S},
  booktitle = {Effective teaching and learning: Perspectives, strategies and implementation},
  pages     = {1--43},
  year      = {2018},
  publisher = {Nova Science Publishers}
}

@ARTICLE{kapur2015learning,
  title     = {Learning from productive failure},
  author    = {Kapur, Manu},
  journal   = {Learning: Research and Practice},
  volume    = {1},
  number    = {1},
  pages     = {51--65},
  year      = {2015},
  publisher = {Taylor \& Francis}
}

@ARTICLE{maier1976learned,
  title     = {Learned helplessness: Theory and evidence},
  author    = {Maier, Steven F and Seligman, Martin E},
  journal   = {Journal of Experimental Psychology: General},
  volume    = {105},
  number    = {1},
  pages     = {3},
  year      = {1976},
  publisher = {American Psychological Association}
}

@article{chrysanthi2026scaffolding,
  title={Scaffolding Generative {AI} as a Tutor: A Quasi-Experimental Study of Learning Outcomes and Motivational, Cognitive and Metacognitive Processes},
  author={Chrysanthi, Melanou and Maik, Beege},
  journal={Education Sciences},
  volume={16},
  number={4},
  pages={651},
  year={2026},
  publisher={MDPI AG}
}

\end{document}